\documentclass[12pt]{article}
\usepackage{a4}
\usepackage{amsmath}
\usepackage{amssymb}
\usepackage{latexsym}
\usepackage{amssymb}
\usepackage{epsfig}
\usepackage{natbib}
\renewcommand{\theequation}{\arabic{section}.\arabic{equation}}

\def \pd{\partial}

\def \tl#1{\overset{\kern 1pt\circ}{#1}}
\def \TL#1{\overset{\kern -3pt \circ}{#1}}
\def \TLL#1{\overset{\kern -7pt \circ}{#1}}

\def \Bbeta{\boldsymbol{\beta}}

\def \Bv{{\boldsymbol{v}}}

\def \LL{{\cal{L}}}

\begin{document}
\title{{\bf The gauge theory of dislocations: 
a uniformly moving screw dislocation}}
\author{
Markus Lazar~\footnote{
{\it E-mail address:} lazar@fkp.tu-darmstadt.de (M.~Lazar).} 
\\ \\
${}^\text{}$ 
        Emmy Noether Research Group,\\
        Department of Physics,\\
        Darmstadt University of Technology,\\
        Hochschulstr. 6,\\      
        D-64289 Darmstadt, Germany
}

\date{\today}    
\maketitle

\begin{abstract}
In this paper we present the equations of motion of a moving screw dislocation
in the framework of the translation gauge theory of dislocations.
In the gauge field theoretical formulation, a dislocation is a massive gauge
field. 
We calculate the gauge field theoretical solutions of a uniformly
moving screw dislocation. We give the subsonic and supersonic
solutions.
Thus, supersonic dislocations are not forbidden from the field theoretical
point of view. 
We show that the elastic divergences at the dislocation core are removed. 
We also discuss the Mach cones produced by supersonic screw dislocations.
\\

\noindent
{\bf Keywords:} dislocation dynamics; gauge theory of dislocations; 
supersonic motion.\\
\end{abstract}

\section{Introduction}

In the dynamics of dislocations it is usually assumed that the screw dislocation 
possesses a Lorentz symmetry~\citep{Frank49, HL, Gunther88}
in contrast to the edge dislocation, which does not
have a Lorentz-type symmetry because of two characteristic velocities, 
namely the velocities of transversal and longitudinal waves, 
entering the field equations. 
By means of such a Lorentz transformation, the screw dislocation can be 
transformed into a steady-state one like in Maxwell's theory of electromagnetic 
fields. 
The question arises if the `classical' dynamics of dislocations reproduces
the correct behavior of dislocations.
Usually, the conventional theories of dislocation dynamics
have several drawbacks (e.g. inertial effects are missing, singularities).  
Unlike special relativity where the speed of light is an upper limit,
in elastodynamics the speed of sound is not a limit velocity. Shock waves
can move faster than the velocity of sound and form Mach cones. 
Similar predictions that dislocations can move faster than the speed of sound
have been given in the literature based on the dynamics of dislocations
(see, e.g.,~\citet{Eshelby56,W67,Gunther68,Gunther69,WW80,XM80,XM08}) and 
computer simulations~\citep{GG99,KK02,Li02,Tsuzuki08}. Supersonic dislocations
have been recently observed in plasma crystals~\citep{Nosenko07}. 

Another question comes up whether
dislocations cause symmetric or asymmetric force stresses.
In classical dislocation theories the elastic distortion tensor is asymmetric
and the force stress tensor is assumed to be symmetric. 
Due to this asymmetry between the 
distortion and the stress tensors, 
\citet{Kroener76} has introduced a modulus of rotation
for the antisymmetric part of the force stress tensor.
In addition, \citet{HK65} and \citet{Kroener68} 
argued that dislocations produce
moment stresses as response and antisymmetric force stresses.
\citet{Pan} calculated the antisymmetric stress corresponding to the body couple
using nonlinear and linear continuum mechanics.
\citet{Gao99} has developed an asymmetric theory of nonlocal elasticity 
based on an atomic lattice model and the three-dimensional  quasicontinuum
field theory. He pointed out that both strain and
local rotation should be regarded as basic variables of geometric deformation
and he has shown that the local rotation makes a very important contribution to
the internal energy. For isotropic materials, the antisymmetric stress comes
from the long range property of interaction of atoms in the metal materials
and non-uniform distribution of atomic forces in the more microscopic structures~\citep{Gao99}.

A promising and straightforward 
candidate for an improved dynamical theory of dislocations 
is the so-called translational gauge theory of
dislocations~\citep{Edelen83,Edelen88,LA08,LA08b}. 
In the gauge theory of defects, dislocations arise naturally as a consequence
of
broken translational symmetry and their existence is not required to be
postulated a priori.
Moreover, such a theory uses the field theoretical framework which is well accepted in 
theoretical physics. 

Recently, \citet{Sharma}  have applied the gauge theory of dislocations to
a uniformly moving screw dislocation. They 
derived a gauge field theoretical solution to the problem
of a uniformly moving screw dislocation. 
However, we will show that their solution is not the correct one for a supersonic
screw dislocation.  
First of all, they used the constitutive relations given
by~\citet{Edelen83,Edelen88}, which are too simple. 
Even in Edelen's version~\citep{Edelen83,Edelen88} 
of the gauge theory of dislocations 
there are at least three characteristic velocities.
For the anti-plane strain problem of a screw dislocation
Edelen's theory possesses two characteristic velocities, 
namely the velocity of shear waves and a gauge theoretical characteristic velocity. 
This makes it impossible without further simplifications and arguments
to use a Lorentz transformation
or at least it is not possible to transform a moving screw dislocation 
into a steady-state dislocation.
Nevertheless, \citet{Sharma} have constructed a solution of a screw dislocation 
using the Lorentz transformation with respect to a gauge theoretical characteristic
velocity. In addition, they have considered the gauge theoretical characteristic
velocity as an upper limit velocity.
Finally, they solved the equations of motion of a screw 
dislocation in a steady state. 
The stress field of their solution does not possess the correct far field
behavior
because their solution is only given in terms of the characteristic gauge
velocity and not in terms of the shear speed of sound.
Neither no Mach cones appear in the supersonic regime.
At best, the solution given by~\citet{Sharma} is only valid 
in the subsonic regime 
if the characteristic gauge velocity is equal to the shear speed of sound.

In the meantime, \citet{LA08,LA08b} have presented an improved 
translational gauge theoretical formulation of dislocations.
They have used the correct isotropic constitutive relations 
in the gauge theory of dislocations which are incomplete and too simple 
in the books of~\citet{Edelen83} and \citet{Edelen88}.
In the present paper we derive the equations of motion of a screw dislocation and we will solve them for a uniformly moving screw dislocation
using the theory of~\citet{LA08}. 
We will show that a dislocation is a massive field in the gauge theoretical
formulation.
We will calculate subsonic and supersonic solutions of a uniformly moving
screw dislocation.

\section{Gauge theory of dislocations}

\setcounter{equation}{0}

In dislocation dynamics we have the following state quantities of 
dislocations\footnote{We use the usual notations: $\beta_{ij,k}:=\pd_k \beta_{ij}$ and 
$\dot{\beta}_{ij}:=\pd_t \beta_{ij}$.} \citep{Kosevich,Landau}
\begin{align}
 \label{disl-den}
T_{ijk}=\beta_{ik,j} - \beta_{ij,k},\qquad    
I_{ij}=-v_{i,j} + {\dot{\beta}}_{ij}
\end{align}  
called the dislocation density tensor and 
the dislocation current tensor, respectively. 
These are the kinematical quantities of dislocations 
and they are given in terms of the incompatible 
elastic distortion tensor $\beta_{ij}$ and incompatible 
physical velocity of the material continuum $v_i$.
The dislocation density tensor describes the location and the shape of 
the dislocation core.
The dislocation current tensor $I_{ij}$ 
describes the movement of the dislocation core and it contains rate terms ($v_{i,j}$ and
$\dot\beta_{ij}$). 
The dislocation density and the dislocation current tensors have to satisfy the translational Bianchi identities (square brackets indicate skewsymmetrization)
\begin{align}
\label{Bianchi-iden}
\epsilon_{jkl}T_{ijk,l}=0,\qquad 
\dot{T}_{ijk} + 2\,I_{i[j,k]}= 0.    
\end{align}
The first equation means that dislocations do not have sources and the 
second one represents that the circulation of the dislocation current is
proportional to the time-derivative of the dislocation density.
On the other hand, (\ref{Bianchi-iden}) are compatibility conditions ensuring that 
$T_{ijk}$ and $I_{ij}$ can be given in terms of $\beta_{ij}$ and $v_i$ according to (\ref{disl-den}).
 
In the dynamical translational gauge theory  of dislocations the Lagrangian
is of the bilinear form (linear theory)
\begin{align}
\label{tot-Lag}
\LL =T-W=
\frac{1}{2}\,p_{i}v_{i} 
+ \frac{1}{2}\,D_{ij}I_{ij}-\frac{1}{2}\,\sigma_{ij}\beta_{ij}  
- \frac{1}{4}\,H_{ijk}T_{ijk}.
\end{align}
Here, the canonical conjugate quantities (response quantities) are defined by
\begin{align}
\label{can-qua}
p_i:=\frac{\pd \LL}{\pd v_i},\qquad
\sigma_{ij}:=-\frac{\pd \LL}{\pd \beta_{ij}} ,\qquad
D_{ij} :=\frac{\pd \LL}{\pd I_{ij}},\qquad
H_{ijk}:=-2\frac{\pd \LL}{\pd T_{ijk}},
\end{align}
where $p_i$, $\sigma_{ij}$, $I_{ij}$, and $H_{ijk}$ are the momentum vector,
the force stress tensor, the dislocation momentum flux tensor, and the pseudomoment stress
tensor, respectively.

The Euler-Lagrange equations derived from the total Lagrangian 
$\LL=\LL(v_i,\beta_{ij},I_{ij},T_{ijk})$ are given by
\begin{align}
\label{euler-lag-2} 
&E^{\, \Bv}_i(\LL)= \pd_t  \frac{\pd \LL}{\pd \dot{v}_i} 
+ \pd_j\frac{\pd \LL}{\pd v_{i,j}} - \frac{\pd \LL}{\pd v_{i}} = 0,\\
\label{euler-lag-3}
&E^{\, \Bbeta}_{ij}(\LL)=\pd_t \frac{\pd \LL}{\pd \dot{\beta}_{ij}} 
+ \pd_k \frac{\pd \LL}{\pd \beta_{ij,k}} - \frac{\pd \LL}{\pd \beta_{ij}}  = 0.
\end{align}
We add to $\LL$ a so-called null Lagrangian, $\LL_N=\sigma^0_{ij}\beta_{ij}-p^0_i v_i$, 
with the ``background'' stress $\sigma^0_{ij}$
and momentum $p^0_i$ as external source fields.
Written in terms of the canonical conjugate quantities~(\ref{can-qua}),
Eqs.~(\ref{euler-lag-2}) and (\ref{euler-lag-3}) take then the form
\begin{alignat}{2}
\label{inhom-di-1}
 D_{ij,j}+ p_i&=p^0_i,
\qquad &&(\text{momentum balance of dislocations})\\
\label{inhom-di-2}
\dot{D}_{ij} + H_{ijk,k}+ \sigma_{ij}&=\sigma^0_{ij}
\qquad &&(\text{stress balance of dislocations}).
\end{alignat}
Eqs.~(\ref{inhom-di-1}) and (\ref{inhom-di-2})
represent the dynamical equations for the balance of dislocations.
Eq.~(\ref{inhom-di-1}) is the momentum balance law of dislocations, where 
the physical momentum is the source of the dislocation momentum flux.
Eq.~(\ref{inhom-di-2}) represents the stress balance of dislocations.
Thus, the force stress and the time derivative of the dislocation momentum flux
are the sources of the pseudomoment stress.
The conservation of linear momentum 
appears as an integrability condition for
the balance of dislocation equations. This can be seen by applying 
$\pd_t$ on~(\ref{inhom-di-1}) and $\pd_j$ on~(\ref{inhom-di-2}) and subtracting 
the last from the first one
\begin{alignat}{2}
\label{inhom-di}
\dot{p}_i -  \sigma_{ij,j}&= 0,
\qquad &&(\text{force balance of elasticity}),
\end{alignat}
where the time-derivative of the physical momentum is the source of the force
stress.

The linear constitutive relations
for the momentum, the asymmetric force stress, the dislocation momentum flux tensor and 
the pseudomoment stress of an isotropic and centrosymmetric medium are~\citep{LA08}
\begin{align}
\label{con-eq}
p_i&=\rho v_i,\\
\label{con-eq-2}
\sigma_{ij}&= \lambda \delta_{ij} \beta_{kk} + \mu (\beta_{ij}+\beta_{ji}) + \gamma (\beta_{ij}-\beta_{ji}),\\
\label{con-eq-3}
D_{ij}&= d_1 \delta_{ij} I_{kk} + d_2 (I_{ij} + I_{ji}) + d_3 (I_{ij} - I_{ji}),\\
\label{con-eq-4}
H_{ijk}&= c_1 T_{ijk} + c_2 (T_{jki} + T_{kij}) + c_3 (\delta_{ij}T_{llk} + \delta_{ik}T_{ljl}),
\end{align}
where $\rho$ is the mass density and with 9 material constants
$\mu$, $\lambda$, $\gamma$, $c_1,\dots,c_3$ and $d_1,\dots,d_3$. 
Here $\mu$, $\lambda$ are the so-called Lam\'e coefficients, $\gamma$ denotes
the modulus of rotation, $c_1,\dots,c_3$ are higher-order stiffness parameters and
$d_1,\dots,d_3$ are related to higher-order inertia due to dislocations.

The requirement of non-negativity of the energy (material stability) $E=T+W\ge~0$ 
leads to the conditions of positive  semi-definiteness of the material constants.
Thus, the material parameters have to fulfill the following conditions~\citep{LA08}
\begin{alignat}{3}
\label{IE-d}
d_2&\ge 0,\qquad &d_3 &\ge 0,\qquad  &3 d_1+2d_2&\ge 0,\\
\label{IE-mu}
\mu&\ge0,\qquad &\gamma&\ge 0,\qquad &3\lambda+2\mu&\ge 0,\\
\label{IE-c}
c_1-c_2&\ge 0,\qquad &c_1+2c_2&\ge 0, \qquad &c_1-c_2+2 c_3&\ge 0.
\end{alignat}

Substituting the constitutive equations in the above 
system~(\ref{inhom-di-1}) and (\ref{inhom-di-2}),
we obtain
\begin{align}
\label{dyn-sys2}
& d_1 ({\dot{\beta}}_{jj,i}-v_{j,ji})
+(d_2+d_3)({\dot{\beta}}_{ij,j}-v_{i,jj})
+(d_2-d_3)({\dot{\beta}}_{ji,j}-v_{j,ji})
+p_i=p^0_i,\\
\label{dyn-sys3}
&d_1\delta_{ij}({\ddot{\beta}}_{kk}-{\dot{v}}_{k,k}) 
+(d_2+d_3)(
{\ddot{\beta}}_{ij}- {\dot{v}}_{i,j})
+(d_2-d_3)({\ddot{\beta}}_{ji}-{\dot{v}}_{j,i})\nonumber\\
&\quad 
+ c_1(\beta_{ik,jk}-\beta_{ij,kk}) 
+c_2(\beta_{ji,kk}-\beta_{jk,ik}+\beta_{kj,ik}-\beta_{ki,jk}) 
+c_3\big[\delta_{ij}(\beta_{lk,lk}-\beta_{ll,kk})\nonumber \\ 
&\quad 
+ (\beta_{kk,ji}-\beta_{kj,ki})\big] 
+\sigma_{ij}
=\sigma^0_{ij},
\end{align} 
which is a coupled system of partial differential equations for the field quantities
$\Bv$ and $\Bbeta$.

\section{Equations of motion of a screw dislocation} 

\setcounter{equation}{0}

We now proceed to derive the equations of motion
for a moving screw dislocation.
The symmetry of a screw dislocation leaves only the following non-vanishing 
components of the physical velocity vector and elastic distortion tensor:
$v_z$, $\beta_{zx}$, $\beta_{zy}$.
The equations of motion of a moving screw dislocation read 
\begin{align}
\label{v-z1}
(d_2+d_3)(\dot{\beta}_{zx,x}+\dot{\beta}_{zy,y}-\Delta v_z)+\rho v_z&=\rho
v^0_z,\\
\label{B-zx1}
(d_2+d_3)(\ddot{\beta}_{zx}-\dot{v}_{z,x})+c_1(\beta_{zy,xy}-\beta_{zx,yy})+(\mu+\gamma)\beta_{zx}&=(\mu+\gamma)\beta^0_{zx},\\
(d_2-d_3)(\ddot{\beta}_{zx}-\dot{v}_{z,x})+c_2(\beta_{zx,yy}-\beta_{zy,xy})+(\mu-\gamma)\beta_{zx}&=(\mu-\gamma)\beta^0_{zx},\\
(d_2+d_3)(\ddot{\beta}_{zy}-\dot{v}_{z,y})+c_1(\beta_{zx,xy}-\beta_{zy,xx})+(\mu+\gamma)\beta_{zy}&=(\mu+\gamma)\beta^0_{zy},\\
\label{B-zy2}
(d_2-d_3)(\ddot{\beta}_{zy}-\dot{v}_{z,y})+c_2(\beta_{zy,xx}-\beta_{zx,xy})+(\mu-\gamma)\beta_{zy}&=(\mu-\gamma)\beta^0_{zy},
\end{align}
where $\Delta=\pd_{xx}+\pd_{yy}$.
In addition, the equilibrium condition is given by
\begin{align}
\label{EC}
(\mu+\gamma)(\beta_{zx,x}+\beta_{zy,y})=\rho \dot{v}_z.
\end{align}
From the system of equations~(\ref{B-zx1})--(\ref{B-zy2}) we obtain the two
relations
\begin{align}
\label{Rel-c}
\frac{c_1}{\mu+\gamma}&=-\frac{c_2}{\mu-\gamma} ,\\
\label{Rel-d}
\frac{d_2+d_3}{\mu+\gamma}&=\frac{d_2-d_3}{\mu-\gamma} .
\end{align}
We may introduce the following quantities
\begin{align}
\ell^2_1&=\frac{c_1}{\mu+\gamma},\\
L^2_1&=\frac{d_2+d_3}{\rho},\\
\tau^2_1&=\frac{d_2+d_3}{\mu+\gamma}.
\end{align}
Here $\ell_1$ and $L_1$ are the `static' and `dynamic'  characteristic length
scales 
and $\tau_1$ is the characteristic time scale of the anti-plane strain problem.
Moreover, $\ell_1$ is related to dislocation stiffness and $L_1$ is related to
dislocation inertia\footnote{We want to mention that similar length scales
($\ell_s$ -- static characteristic length, $\ell_d$  -- dynamic characteristic length)
have been also obtained by~\citet{Aifantis05,Aifantis06,Aifantis07} 
in a dynamic gradient elasticity.}.
The velocity of elastic shear waves is defined in terms of the `dynamic'
length scale $L_1$ and the time scale $\tau_1$:
\begin{align}
\label{cT}
c^2_T=\frac{L_1^2}{\tau_1^2}=\frac{\mu+\gamma}{\rho}.
\end{align}
Due to the presence of $\gamma$ the velocity of elastic shear waves is greater
than in a theory with symmetric force stresses.
Moreover, the velocity of shear waves has a similar form in micropolar 
elasticity (see, e.g.,~\citet{Nowacki}) where the force stress is also asymmetric.
In a similar way, we may introduce the following transversal gauge-theoretical 
velocity defined in terms of 
$\ell_1$ and $\tau_1$:
\begin{align}
\label{aT}
a^2_T=\frac{\ell_1^2}{\tau_1^2}=\frac{c_1}{d_2+d_3},
\end{align}
and therefore
\begin{align}
\label{Rel-a}
\frac{\ell_1^2}{L_1^2}=\frac{a_T^2}{c_T^2}.
\end{align}

In the case $\gamma=0$ we recover symmetric force stresses 
and from relations~(\ref{Rel-c}) and (\ref{Rel-d}): $c_1=-c_2$ and $d_3=0$.
It can be seen that $c_1=-c_2$ gives the inequalities~(\ref{IE-c}) with 
$c_1\ge 0$ and $-c_1\ge 0$.
Thus, for nonnegative $c_1$ and $c_1\neq 0$ we cannot fulfill~(\ref{IE-c}).
This is the price we have to pay for symmetric force stresses of a screw 
dislocation in the dislocation gauge theory.

Because we deal with the physical state quantities $(\Bv,\Bbeta)$,
no pseudo-Lorentz gauge~\citep{Edelen83,Edelen88} is used and allowed 
during the simplification of the equations of motion. Gauge conditions are only
allowed for gauge potentials and not for physical state quantities.
Of course, for anti-plane strain the equilibrium condition~(\ref{EC}) together 
with (\ref{cT}) looks like
a `gauge' condition but it is not a gauge condition 
from the physical interpretation.

Applying the equilibrium condition~(\ref{EC}), the equations of 
motion~(\ref{v-z1})--(\ref{B-zy2})
can be written in the form
\begin{align}
\label{EOM-v}
\tau_1^2\ddot{v}_{z}-L_1^2\Delta v_z+v_z&=v^0_z,\\
\label{EOM-Bzx}
\tau_1^2\ddot{\beta}_{zx}-\ell_1^2 \Delta \beta_{zx} 
-\tau_1^2 \Big(1-\frac{\ell_1^2}{L_1^2}\Big) \dot{v}_{z,x}
+\beta_{zx}&=\beta^0_{zx},\\
\label{EOM-Bzy}
\tau_1^2\ddot{\beta}_{zy}-\ell_1^2 \Delta \beta_{zy} 
-\tau_1^2\Big(1-\frac{\ell_1^2}{L_1^2}\Big) \dot{v}_{z,y}
+\beta_{zy}&=\beta^0_{zy}.
\end{align}
These are the equations of motion of an arbitrary moving screw dislocation
in the framework of dislocation gauge theory.
Some important questions arise if the lengths $\ell_1$ and $L_1$ are 
independent or not and is there a physical reason to decouple the field
equations~(\ref{EOM-v})--(\ref{EOM-Bzy}).

Consider a screw dislocation moving in the $x$-direction.
If we want to construct a solution which is consistent with the classical
solution, we have to fulfill the condition: 
$I_{zx}^0=-v^0_{z,x}+\dot{\beta}^0_{zx}=0$, because only the 
classical dislocation density $T^0_{zxy}$ and dislocation current $I^0_{zy}$
are non-zero.
If we use the field equations~(\ref{EOM-v}) and (\ref{EOM-Bzx}),
we obtain:
\begin{align}
\label{Cond-I}
-\Big[\frac{\ell_1^2}{c_T^2}\, \pd_{tt}-L_1^2\Delta+1\Big]v_{z,x}
+\Big[\frac{L_1^2}{c_T^2}\, \pd_{tt}-\ell_1^2\Delta+1\Big]\dot{\beta}_{zx}=0.
\end{align}
To guarantee that~(\ref{Cond-I}) is fulfilled, we choose
\begin{align}
\label{Rel-L}
v_{z,x}=\dot{\beta}_{zx},\qquad{\text{and}}\qquad
L_1=\ell_1 .
\end{align}
Equation~(\ref{Rel-L}a) ensures that $I_{zx}=0$ and 
(\ref{Rel-L}b) with (\ref{Rel-a}) gives the relation $a_T=c_T$, which means
that only one characteristic velocity $c_T$ survives for a screw dislocation in
the gauge theory of dislocations.
Thus, for a physically consistent solution 
we obtain the uncoupled Klein-Gordon equations
\begin{align}
\label{KGE-v-2}
&\big [1+\ell_1^2 \square_{T} \big] v_z=v^0_z,\\
\label{KGE-Bzx-2}
&\big [1+\ell_1^2 \square_{T} \big] \beta_{zx} 
=\beta^0_{zx},\\
\label{KGE-Bzy-2}
&\big [1+\ell_1^2 \square_{T} \big] \beta_{zy}
=\beta^0_{zy},
\end{align}
with the following $(1+2)$-dimensional d'Alembert operator (wave operator)
\begin{align}
\square_{T}&=\frac{1}{c^2_T}\, \pd_{tt}-\Delta.
\end{align}
Moreover,
the uncoupled system ~(\ref{KGE-v-2})--(\ref{KGE-Bzy-2}) 
with only one characteristic velocity possesses a Lorentz symmetry.
However, $c_T$ is not a limit velocity unlike the speed of light in special 
relativity. 
In field theories, Klein-Gordon equations describe massive fields 
(see, e.g.,~\cite{Rubakov}).
Thus, a dislocation is a massive gauge field.
From the condition $L_1=\ell_1$, we find for the inertia term of a screw
dislocation
\begin{align}
d_2+d_3=\frac{c_1}{c^2_T}=\rho\, \ell^2_1,
\end{align} 
that it is given in terms of the characteristic length scale $\ell_1$. 
Under these assumptions, the dynamical dislocation gauge theory of 
a screw dislocation possesses only
one internal length scale $\ell_1$.
 
If we multiply Eqs.~(\ref{KGE-v-2})--(\ref{KGE-Bzy-2}) with $\square_{T}$ 
and using the `classical' result~\citep{Gunther68,Gunther69}, 
we obtain
\begin{align}
\label{KGE-v-3}
&\big [1+\ell_1^2 \square_{T} \big]\square_{T} v_z=I^0_{zy,y},\\
\label{KGE-Bzx-3}
&\big [1+\ell_1^2 \square_{T} \big]\square_{T} \beta_{zx} 
=T^0_{zxy,y},\\
\label{KGE-Bzy-3}
&\big [1+\ell_1^2 \square_{T} \big] \square_{T}\beta_{zy}
=-T^0_{zxy,x}+\frac{1}{c_T^2} \dot{I}^0_{zy},
\end{align}
as a set of fourth-order partial differential equations. 
As source terms only the classical dislocation density $T^0_{zxy}$ and 
dislocation current $I^0_{zy}$ are acting.
Eqs.~(\ref{KGE-v-3})--(\ref{KGE-Bzy-3}) have the two-dimensional 
form of Bopp-Podolsky equations~\citep{Bopp,Podolsky} (see also~\citet{Iwan}) 
in generalized electrodynamics, introduced by Bopp and Podolsky in order 
to avoid singularities in electrodynamics.

In general, the velocity $V$ of a screw dislocation might be subsonic or
supersonic relative to the material space.
A subsonic velocity lies in the range: $0<V<c_T$ 
and for supersonic screw dislocations the velocity reads: $c_T<V$.

\section{Uniformly moving screw dislocation }
\setcounter{equation}{0}

We now study a screw dislocation moving with a uniform velocity $V$ in
the $x$-direction. Here $V$ denotes the dislocation velocity in the material 
space.
The dislocation moves relative to the material space, which plays the role of
an `aether'.
If a screw dislocation is moving with the velocity $V$, 
then $v_z$ is the physical velocity of the material space, 
where the dislocation lives in, in order 
to transport the dislocation core to another position in the material space. 
Let $x'$ be the coordinate in the direction of motion in the moving coordinate system.
For the uniformly moving screw dislocation, then we use the transform
\begin{align}
x'=x-Vt 
\end{align}
with
\begin{align}
\pd_{t} =-V\pd_{x'}
\end{align}
and we obtain the equations
\begin{align}
\label{v-B}
&\big[1 -\ell_1^2 \big((1-M^2_{T}) \pd_{x'x'}+\pd_{yy}\big)\big] v_z=v^0_z,\\
\label{Bzx-B}
&\big[1 -\ell_1^2\big( (1-M^2_{T}) \pd_{x'x'}+\pd_{yy})\big]\beta_{zx} 
=\beta^0_{zx},\\
\label{Bzy-B}
&\big[1 -\ell_1^2 \big((1-M^2_{T}) \pd_{x'x'}+\pd_{yy}\big)\big]
\beta_{zy}
=\beta^0_{zy},
\end{align}
and
\begin{align}
\label{T-B}
&\big[1 -\ell_1^2\big( (1-M^2_{T}) \pd_{x'x'}+\pd_{yy}\big)\big]
T_{zxy} =T^0_{zxy},\\
\label{Izy-B}
&\big[1 -\ell_1^2\big((1-M^2_{T})  \pd_{x'x'}+\pd_{yy}\big)\big]
I_{zy} =I^0_{zy},
\end{align}
with the Mach number of a moving screw dislocation relative to $c_T$:
\begin{align}
\label{MN}
M_{T}=\frac{V}{c_T}.
\end{align}
Eqs.~(\ref{v-B})--(\ref{Izy-B}) are two-dimensional 
modified inhomogeneous Helmholtz equations 
(subsonic case). In the supersonic case they change to
one-dimensional inhomogeneous Klein-Gordon equations (see below).

\subsection{Subsonic case}

We now study a screw dislocation moving with a uniform velocity $V<c_T$ in
the $x$-direction ($M_{T}<1$).
The `background' solution reads (see, e.g.,~\citet{Gunther68})
\begin{align}
\label{Cl}
&v^0_z= V \pd_y F^0,\qquad
\beta^0_{zx}=-  \pd_y F^0,\qquad
\beta^0_{zy}= \beta_{T}^2 \pd_x F^0,\nonumber\\
&T^0_{zxy}=b \delta(x-Vt)\delta(y),\qquad
I^0_{zy}=-b V \delta(x-Vt)\delta(y),
\end{align}
with
\begin{align}
F^0=\frac{b}{2\pi} \frac{1}{\beta_{T}}\, \ln r_{T}
\end{align}
and
\begin{align}
r_{T}&=\sqrt{(x-Vt)^2 +\beta_{T}^2 y^2},\qquad
\beta_{T}=\sqrt{1-M^2_{T}}.
\end{align}

\begin{figure}[t]\unitlength1cm
\centerline{
(a)
\begin{picture}(7,7)
\put(0.0,0.2){\epsfig{file=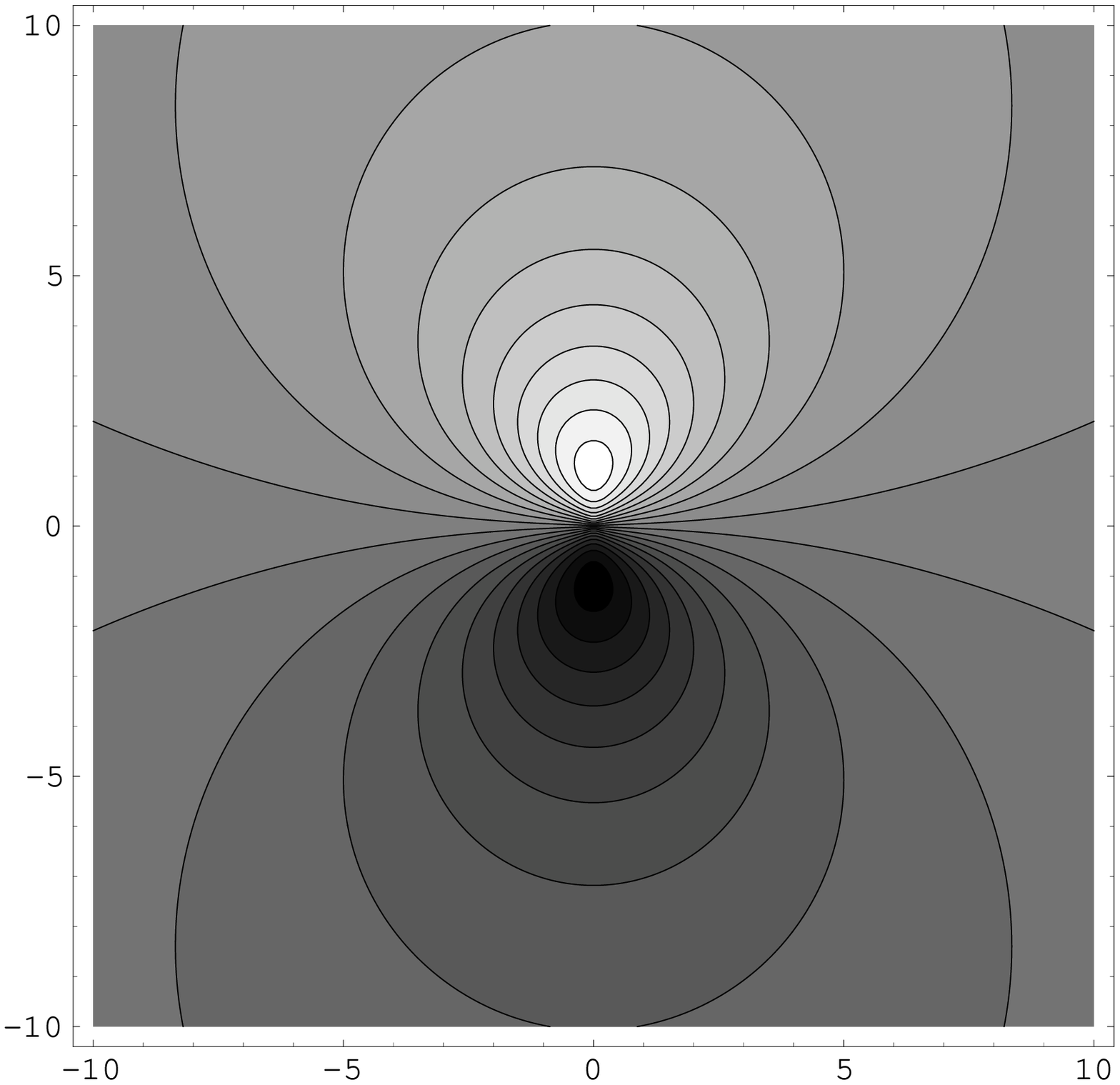,width=7cm}}
\put(3.5,-0.1){$x-Vt$}
\put(-0.5,3.7){$y$}
\end{picture}
(b)
\begin{picture}(7,7)
\put(0.0,0.2){\epsfig{file=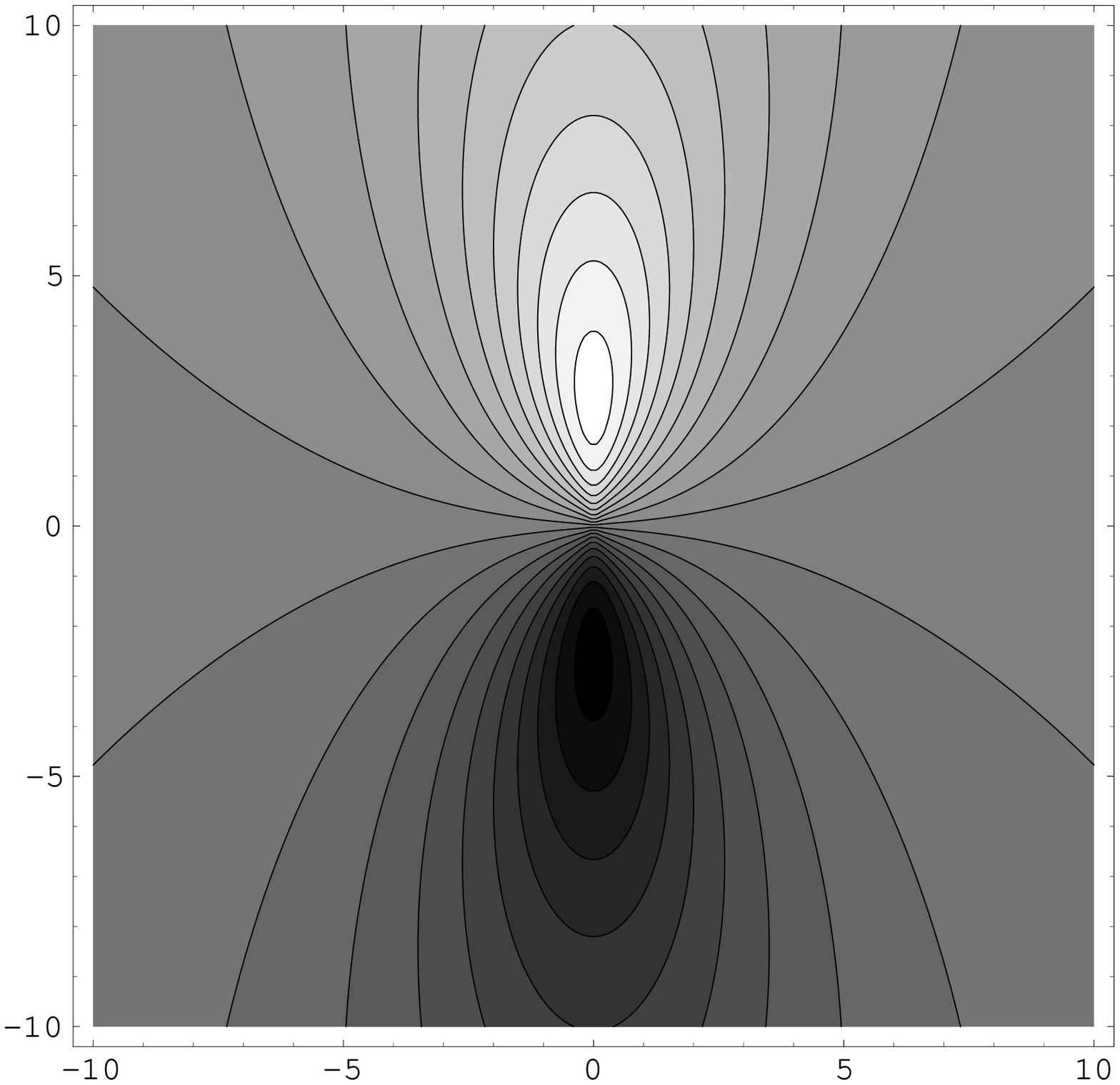,width=7cm}}
\put(3.5,-0.1){$x-Vt$}
\put(-2.0,3.7){$$}
\end{picture}
}
\caption{Contourplot of the elastic velocity $v_z$ of a screw dislocation 
moving with subsonic speed: (a) $V=0.1 c_T$, (b) $V=0.9 c_T$.}
\label{fig:vz}
\end{figure}

\begin{figure}[t]\unitlength1cm
\centerline{
\epsfig{figure=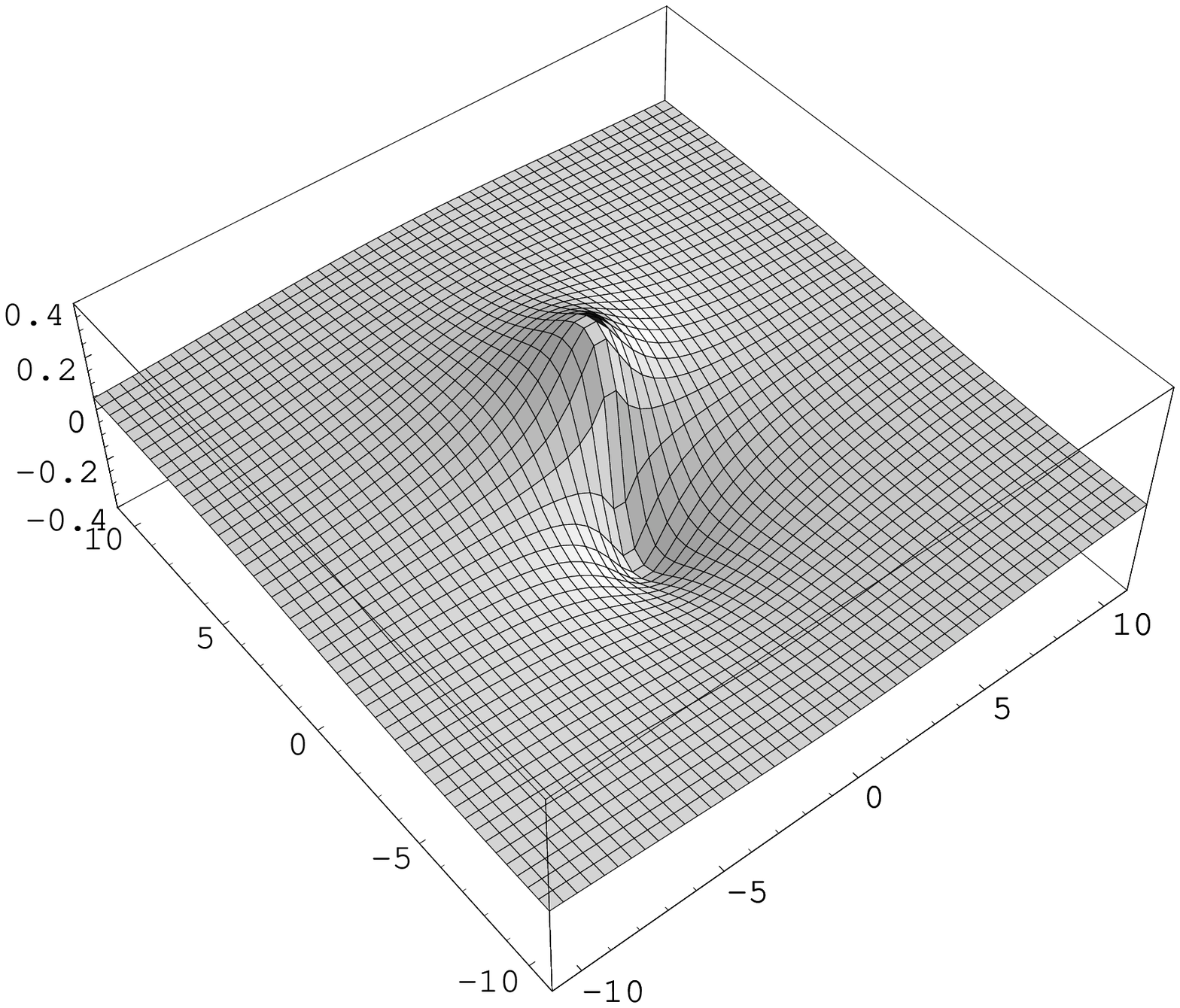,width=7.0cm}
\put(-1.5,1.0){$\frac{x-Vt}{\ell_1\beta_{T}}$}
\put(-6.5,1.0){$\frac{y}{\ell_1\beta_{T}}$}
\put(-6.2,-0.3){$\text{(a)}$}
\hspace*{0.4cm}
\put(0,-0.3){$\text{(b)}$}
\put(5.5,1.0){$\frac{x-Vt}{\ell_1\beta_{T}}$}
\put(0.5,1.0){$\frac{y}{\ell_1\beta_{T}}$}
\epsfig{figure=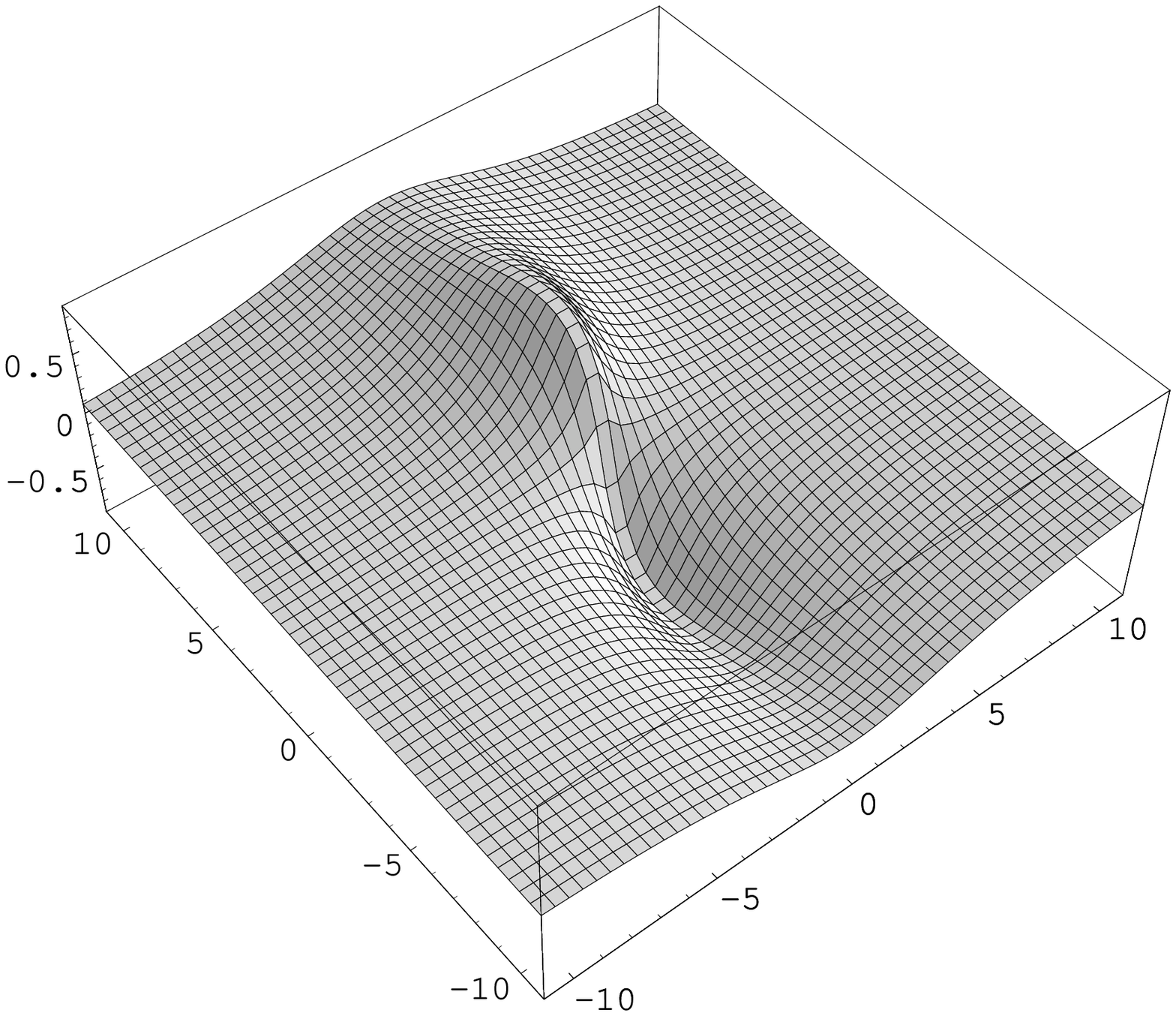,width=7.0cm}
}
\caption{Three-dimensional plots of the physical velocity $v_z$ in units of 
 $bV/2\pi \ell_1^2\beta_T$:
(a)~$V=0.1c_T$,
(b) $V=0.9c_T$.}
\label{fig:vz-3D}
\end{figure}

In the subsonic region, the field equations~(\ref{v-B})--(\ref{Izy-B})
are:
\begin{align}
\label{v-B2}
&\big[1 -\ell_1^2( \beta_{T}^2 \pd_{x'x'}+\pd_{yy})\big] v_{z}
=v^0_z,\\ 
\label{Bzx-B2}
&\big[1 -\ell_1^2( \beta_{T}^2 \pd_{x'x'}+\pd_{yy})\big]\beta_{zx} 
=\beta^0_{zx},\\
\label{Bzy-B2}
&\big[1 -\ell_1^2 (\beta_{T}^2 \pd_{x'x'}+\pd_{yy})\big]
\beta_{zy}
=\beta^0_{zy},
\end{align}
and 
\begin{align}
\label{T-B2}
&\big[1 -\ell_1^2\big( \beta^2_{T} \pd_{x'x'}+\pd_{yy}\big)\big]
T_{zxy} =T^0_{zxy},\\
\label{Izy-B2}
&\big[1 -\ell_1^2\big(\beta^2_{T}  \pd_{x'x'}+\pd_{yy}\big)\big]
I_{zy} =I^0_{zy}.
\end{align}
Eqs.~(\ref{v-B2})--(\ref{Izy-B2}) can be easily
solved with the inhomogeneous parts~(\ref{Cl}) using Fourier
transform or other techniques.
The solutions for the dislocation density and the dislocation flux of a screw
dislocation are given by
\begin{align}
\label{T}
T_{zxy}&=\frac{b}{2\pi} \frac{1}{\ell^2_1 \beta_{T}}\,  K_0\Big(\frac{r_{T}}{\ell_1\beta_{T}}\Big),\\
\label{I}
I_{zy}&=-\frac{b}{2\pi} \frac{V }{\ell^2_1 \beta_{T}}\,  K_0\Big(\frac{r_{T}}{\ell_1\beta_{T}}\Big).
\end{align}
\begin{figure}[t]\unitlength1cm
\centerline{
(a)
\begin{picture}(7,7)
\put(0.0,0.2){\epsfig{file=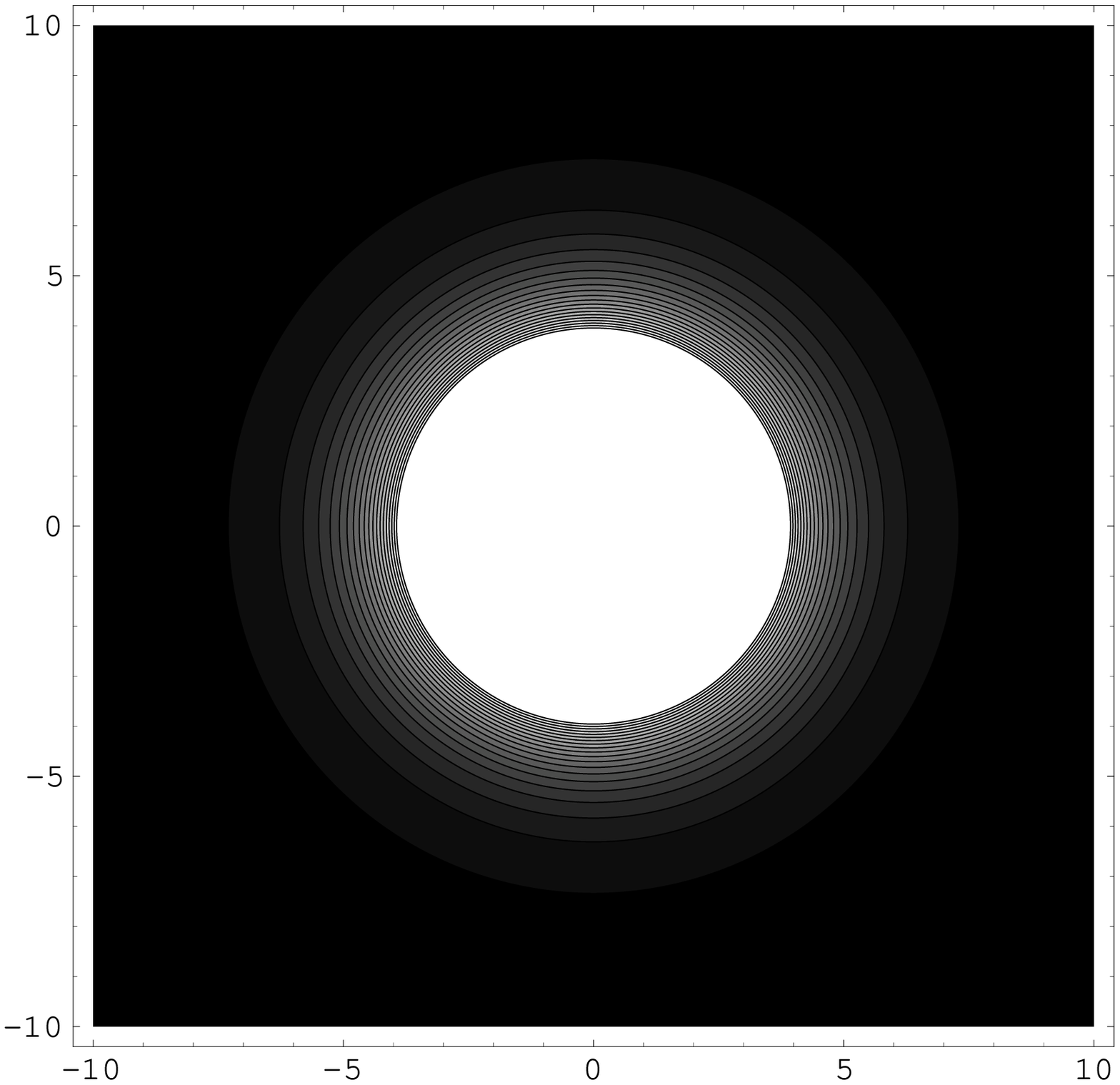,width=7cm}}
\put(3.5,-0.1){$x-Vt$}
\put(-0.5,3.7){$y$}
\end{picture}
(b)
\begin{picture}(7,7)
\put(0.0,0.2){\epsfig{file=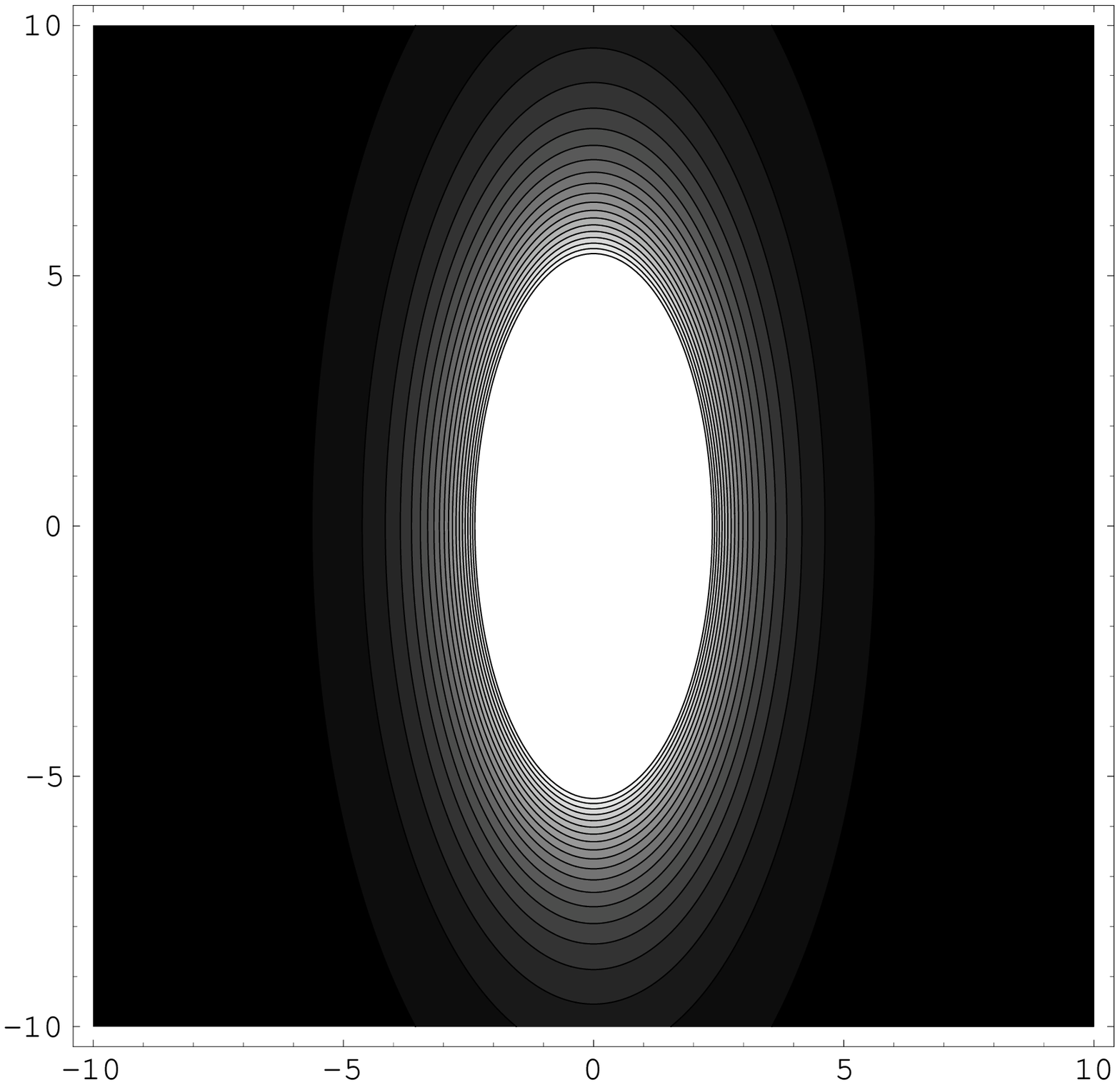,width=7cm}}
\put(3.5,-0.1){$x-Vt$}
\put(-2.0,3.7){$$}
\end{picture}
}
\caption{Contourplot of the dislocation density $T_{zxy}$ of a screw dislocation 
moving with subsonic speed: (a) $V=0.1 c_T$, (b) $V=0.9 c_T$.}
\label{fig:Tzxy}
\end{figure}
Here, the symbol $K_n$ stands for the modified Bessel function of second kind
(McDonald function) and of order $n$. 
In Fig.~\ref{fig:Tzxy} the dislocation density is plotted for different speeds subsonic with respect 
to $c_T$. The field of the dislocation density suffers a
contraction when the value of its velocity approaches the velocity $c_T$. 
The curve of $T_{zxy}$ is a circle at $V=0$ and is a generalized ellipse at any
other velocity $V<c_T$.
The field is dilated in the directions orthogonal to the direction of motion 
and contracted along the line of motion.
The solution for the elastic velocity is given by
\begin{align}
v_z&=\frac{b}{2\pi} \frac{\beta_{T} V y}{r_{T}^2} 
\Big[1-\frac{r_{T}}{\ell_1 \beta_{T}}\,  
K_1\Big(\frac{r_{T}}{\ell_1\beta_{T} }\Big)\Big].
\end{align}
In Figs.~\ref{fig:vz} and \ref{fig:vz-3D} 
the physical velocity of a screw dislocations 
is plotted for subsonic velocities. It shows again a Fitzgerald contraction.
It does not have a singularity. The physical velocity possesses extremum values depending on the dislocation velocity  near the dislocation line. 
The solution of the elastic distortion reads
\begin{align}
\label{Bzx}
\beta_{zx}&=-\frac{b}{2\pi} \frac{\beta_{T} y}{r_{T}^2} 
\Big[1-\frac{r_{T}}{\ell_1 \beta_{T}}\,  
K_1\Big(\frac{r_{T}}{\ell_1\beta_{T} }\Big)\Big],\\
\label{Bzy}
\beta_{zy}&=\frac{b}{2\pi} \frac{\beta_{T} (x-Vt)}{r_{T}^2} 
\Big[1-\frac{r_{T}}{\ell_1 \beta_{T}}\,  
K_1\Big(\frac{r_{T}}{\ell_1\beta_{T} }\Big)\Big].
\end{align}
All the elastic fields are nonsingular in the gauge theoretical framework.

On the other hand, the solution~(\ref{Bzx}) and (\ref{Bzy})
has the same form as the expression given by~\citet{Sharma} if $c_T=a_T$ 
and $\gamma=0$.

\subsection{Supersonic case}
\begin{figure}[tp]\unitlength1cm
\centerline{
\begin{picture}(7,7)
\put(0.0,0.2){\epsfig{file=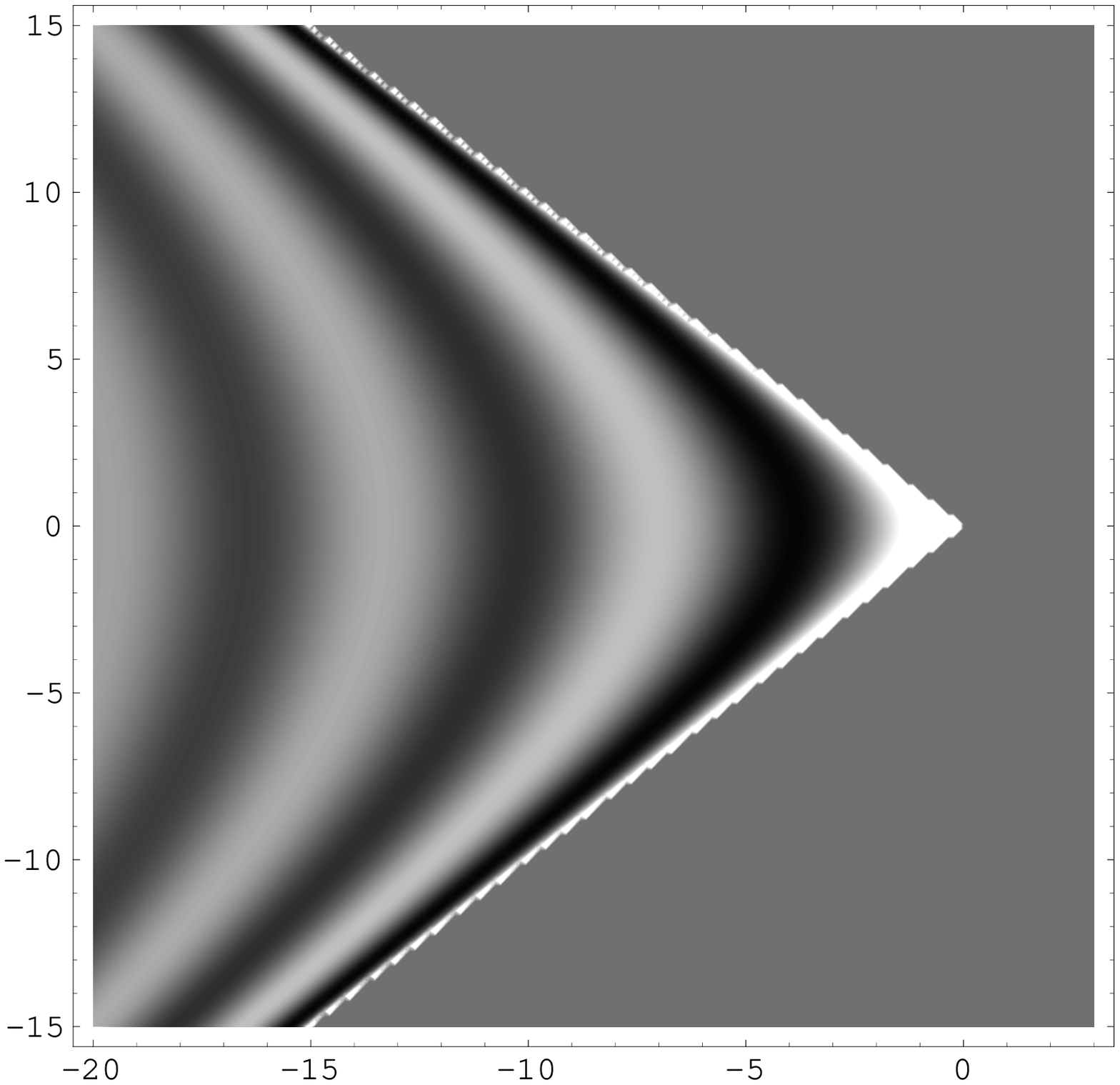,width=7cm}}
\put(3.5,-0.1){$x-Vt$}
\put(-0.5,3.7){$y$}
\end{picture}
}
\caption{Contourplot of the dislocation density $T_{zxy}$ of a screw dislocation 
moving with supersonic speed $V=\sqrt{2} c_T$.}
\label{fig:T-SS}
\end{figure}

We now consider the supersonic case: $V>c_T$, 
(the supersonic case: $M_{T}>1$).
Therefore, the term $\beta^2_{T}=-\gamma^2_{T}$ alters where
\begin{align}
\gamma_{T}=\sqrt{M^2_{T}-1}.
\end{align}
Then the field equation for the dislocation density changes to
\begin{align}
\label{T-B-SS}
&\big[1 +\ell_1^2( \gamma_{T}^2 \pd_{x'x'}-\pd_{yy})\big]
T_{zxy} =T^0_{zxy}.
\end{align}
It is an inhomogeneous one-dimensional Klein-Gordon equation.
The dislocation density of a supersonic screw dislocation has
the form of the corresponding Green function. Therefore, it is given by 
(see, e.g., \citet{Iwan})
\begin{align}
\label{T-SS}
T_{zxy}=\frac{b}{2\ell_1^2\gamma_{T}}\, 
J_0\bigg(\frac{\sqrt{(Vt-x)^2-\gamma_{T}^2 y^2}}{\ell_1 \gamma_{T}}\bigg)
\, 
H((Vt-x) -\gamma_{T} |y|).
\end{align}
Here, the symbol $J_n$ denotes the Bessel function of first kind and of order $n$
and $H$ is the Heaviside step function.
It is non-zero just for $Vt-x>\gamma_{T}|y|$.
It builds a shear-wave Mach cone with the angle $\sin \theta_{T} =\frac{c_T}{V}$.
The dislocation density has a maximum of $T_{zxy}=b/[2\gamma_{T} \ell_1^2]$ 
on the Mach cone. Inside the Mach cone it oscillates with decreasing amplitude.
For the dislocation current density we obtain $I_{zy}=-V\, T_{zxy}$ with (\ref{T-SS}).
The visualization of the Mach cone of the dislocation density $T_{zxy}$ produced
by the screw dislocation in the supersonic regime is plotted in
Fig.~\ref{fig:T-SS}. 

\begin{figure}[tp]\unitlength1cm
\centerline{
\begin{picture}(7,7)
\put(0.0,0.2){\epsfig{file=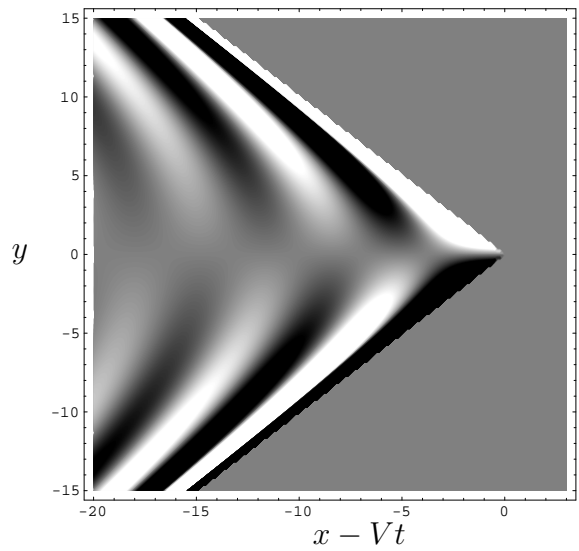,width=7cm}}
\put(3.5,-0.1){$x-Vt$}
\put(-0.5,3.7){$y$}
\end{picture}
}
\caption{Contourplot of the physical velocity $v_z$ of a screw dislocation 
moving with supersonic speed $V=\sqrt{2} c_T$.}
\label{fig:v-SS}
\end{figure}

The equations of the elastic fields, altering their character from elliptic to hyperbolic, are
\begin{align}
&\big[1+\ell_1^2 (\gamma_{T}^2\pd_{x'x'}- \pd_{yy})\big]v_z =v_z^0,\\
&\big[1+\ell_1^2 (\gamma_{T}^2\pd_{x'x'}- \pd_{yy})\big]\beta_{zx} =\beta_{zx}^0,\\
\label{Bzy-B21-SS}
&\big[1 +\ell_1^2 (\gamma_{T}^2 \pd_{x'x'}-\pd_{yy})\big]\beta_{zy}
=\beta^0_{zy},
\end{align}
with the inhomogeneous parts (\ref{vz-SS-0-S})--(\ref{Bzy-SS-0-S}).
The corresponding solutions read
\begin{align}
\label{v-TS}
v_{z}&=\frac{b V}{2 \ell_1}\, \frac{y}{\sqrt{(Vt-x)^2-\gamma_{T}^2 y^2}}\,
J_1\bigg(\frac{\sqrt{(Vt-x)^2-\gamma_{T}^2 y^2}}{\ell_1 \gamma_{T}}\bigg)
\, 
H((Vt-x) -\gamma_{T} |y|),\\
\label{Bzx-TS}
\beta_{zx}&=-\frac{b }{2\ell_1}\, \frac{y}{\sqrt{(Vt-x)^2-\gamma_{T}^2 y^2}}\,
J_1\bigg(\frac{\sqrt{(Vt-x)^2-\gamma_{T}^2 y^2}}{\ell_1 \gamma_{T}}\bigg)
\, 
H((Vt-x) -\gamma_{T} |y|),\\
\label{Bzy-TS}
\beta_{zy}&=-\frac{b }{2\ell_1}\, \frac{(Vt-x)}{\sqrt{(Vt-x)^2-\gamma_{T}^2 y^2}}\,
J_1\bigg(\frac{\sqrt{(Vt-x)^2-\gamma_{T}^2 y^2}}{\ell_1 \gamma_{T}}\bigg)
\, 
H((Vt-x) -\gamma_{T} |y|).
\end{align}
It can be seen that (\ref{v-TS}) is non-zero just for $Vt-x>\gamma_{T}|y|$.
Also it builds a shear-wave Mach cone with the angle $\sin \theta_{T} =\frac{c_T}{V}$.
It is important to note that the classical singularity as Dirac delta function
of supersonic dislocations on the Mach cone does not appear in our gauge
theoretical result. The elastic distortions and the velocity have a maximum
value on the Mach cone: $v_z=bVy/[4 L_1^2 \gamma_{T}]$,
$\beta_{zx}=-b y/[4\ell_1^2\gamma_{T}]$,
$v_z$ and $\beta_{zx}$ are also zero at $y=0$.
Inside the Mach cone they oscillate with decreasing amplitude.
The visualization of the Mach cone of the physical velocity $v_z$ produced
by the screw dislocation in the supersonic regime is plotted in Fig.~\ref{fig:v-SS}.

Let us mention that the solution of~\citet{Sharma} does not possess the correct
behavior in the supersonic region.
The solution given by~\citet{Sharma} does not show  Mach cones,
which are predicted in computer simulations~\citep{GG99,KK02,Li02,Tsuzuki08} and found experimentally~\citep{Nosenko07}.


\section{Conclusion}
In this paper, we have investigated a moving screw dislocation in 
the gauge field theory of dislocations. 
We derived the equations of motion of an arbitrary moving screw dislocation 
in this field-theoretical framework. 
First, a coupled system of inhomogeneous Klein-Gordon equations is obtained
in the dynamical case with two characteristic velocities.
We have found one dynamical characteristic length scale $L_1$, one static 
characteristic length scale $\ell_1$ and one characteristic time scale $\tau_1$.
Later we have decoupled the field equations, following physical arguments, to
construct a consistent solution. 
Due to these arguments we found $L_1=\ell_1$ and $a_T=c_T$.
The elastic fields $v_z$, $\beta_{zx}$ and $\beta_{zy}$ 
and the fields of the dislocation core $T_{zxy}$ and $I_{zy}$ 
have the characteristic speed $c_T$. 
For a uniformly moving screw dislocation we have given analytical solutions
for the subsonic and supersonic cases.
For the supersonic case we found one Mach cone for the velocities $c_T$.

\section*{Acknowledgement}
The author has been supported by an Emmy-Noether grant of the 
Deutsche Forschungsgemeinschaft (Grant No. La1974/1-2). 

\begin{appendix}
\setcounter{equation}{0}
\renewcommand{\theequation}{\thesection.\arabic{equation}}
\section{Supersonic screw dislocation in elasticity}
\label{appendixA}
In the elasticity, a supersonic screw dislocation moves with
the velocity: $V>c_T$. 
In symmetric elasticity the speed of shear waves reads: $c^2_T=\mu/\rho$ and
in asymmetric elasticity it is given by $c^2_T=(\mu+\gamma)/\rho$.
In asymmetric elasticity the antisymmetric part of the stress tensor produces
body couples.
 The field equations for the elastic velocity and 
the elastic distortions read~\citep{Gunther68}
\begin{align}
\label{v-SS-0}
&\square_{T} v_z^0=-b V\, \pd_y \delta(x-Vt)\delta(y),\\
\label{Bzx-SS-0}
&\square_{T} \beta_{zx}^0=b\, \pd_y \delta(x-Vt)\delta(y),\\
\label{Bzy-SS-0}
&\square_{T} \beta_{zy}^0=b\, \gamma^2_{T}\pd_{x} \delta(x-Vt)\delta(y).
\end{align}
These are inhomogeneous wave equations describing massless fields.
The supersonic solutions are
\begin{align}
\label{vz-SS-0-S}
v_z^0&=\frac{bV}{2}\frac{y}{|y|}\, \delta(Vt-x-\gamma_{T}|y|),\\
\beta_{zx}^0&=-\frac{b}{2}\frac{y}{|y|}\, \delta(Vt-x-\gamma_{T}|y|),\\
\label{Bzy-SS-0-S}
\beta_{zy}^0&=-\frac{b\,\gamma_{T}}{2}\, \delta(Vt-x-\gamma_{T}|y|),
\end{align}
where $\delta$ denotes the Dirac delta function.
Thus, the classical solutions are shock waves produced by a supersonic screw dislocation. Both in front of and behind the shock front the material is undeformed 
and at rest. 

\end{appendix}

\end{document}